\documentclass[11pt] {article} 

\usepackage{amsmath,amssymb}

\usepackage[all]{xy}
\usepackage{color}

\usepackage[ansinew]{inputenc}
\usepackage[T1]{fontenc}
\usepackage{lmodern}
\usepackage{setspace}

\begin{document}

\title{The Hilbert space of conditional clauses}

\author{Charles Francis\thanks{Jesus College, Cambridge; e-mail: C.E.H.Francis.75@cantab.net }}


\maketitle

\begin{abstract} 
In the absence of a satisfactory interpretation of quantum theory, physical law lacks physical basis. This paper reviews the orthodox, or Dirac-von Neumann interpretation, and makes explicit that Hilbert space describes propositions about measurement results. Kets are defined as conditional clauses referring to measurements in a formal language. It is seen that these clauses are elements of a Hilbert space, such that addition is logical disjunction, the dual space consists of consequent clauses, and the inner product is a set of statements in the subjunctive mood. The probability interpretation gives truth values for corresponding future tense statements when the initial state is actually prepared and the final state is to be measured. The mathematical structure of quantum mechanics is formulated in terms of discrete measurement results at finite level of accuracy and does not depend on an assumption of a substantive, or background, space-time continuum. A continuum of kets, $ \left|x\right\rangle$ for $x\in \mathbb{R}^{3}$, is constructed from linear combinations of kets in a finite basis. The inner product can be expressed either as a finite sum or as an integral. Discrete position functions are uniquely embedded into smooth wave functions in such a way that differential operators are defined. It is shown that the choice of basis has no effect on underlying physics (quantum covariance). The Dirac delta has a representation as a smooth function. Operators do not in general have an integral form. The Schr\"{o}dinger equation is shown from the requirements of the probability interpretation. It is remarked that a formal construction of qed avoiding divergence problems has been completed using finite dimensional Hilbert space. I conclude that quantum mechanics makes statements about the world with clear physical meaning, such that space is emergent from particle interactions and has no fundamental role. 
\end{abstract}

\begin{footnotesize}
 

\end{footnotesize}
\newpage

\section{Introduction}
\label{sec:Introduction}
\subsection{Objectives}
\label{Objectives}
In a review of \textit{Foundations of quantum physics} by C. Piron, V. S. Varadarajan \cite{Varadarajan} remarked \textit{``While an `explanation' of the axioms is nowadays regarded as unnecessary in a mathematical treatise, it is still an important part in any exposition of the mathematical nature of physical theories''}. It has been a major issue that such an explanation has been lacking in quantum theory, usually only resolved by adopting a \textit{philosophy} that physics also requires no explanations. It would be far better to resolve this issue by providing an explanation. To do so, it is not sufficient simply to \textit{assume} the axiomatic structure of Hilbert space or some equivalent mathematical structure. Rather, we must \textit{exhibit} something with the properties of Hilbert space. In this paper, superposition will be exhibited as weighted logical \textsc{or} applied to conditional and consequent clauses in a formal language describing possible measurement results, and the inner product will be identified with sentences in the subjunctive mood constructed from these conditional and consequent clauses.

Carlo Rovelli \cite{Rovelli1}  
describes the purpose of Relational Quantum Mechanics: \textit{``\ldots to do for the formalism of quantum mechanics what Einstein did for the Lorentz transformations: i. Find a set of simple assertions about the world, with clear physical meaning, that we know are experimentally true (postulates); ii. Analyze these postulates, and show that from their conjunction it follows that certain common assumptions about the world are incorrect; iii. Derive the full formalism of quantum mechanics from these postulates. I expect that if this program could be completed, we would at long last begin to agree that we have understood quantum mechanics''}. 

To say that we have completed such a program it is not sufficient to present a consistent mathematical structure giving correct predictions. A mathematical model is defined from its axioms. In physics we should require that the axioms are physically sensible in addition to being logically consistent and empirically true. The defining axioms for the mathematical structure described here will be termed postulates and definitions; postulates are intended to contain empirical assertions about the world, while definitions are purely semantic. There is some subjectivity in assessing whether a definition should be termed a postulate, but this does not affect mathematical structure. In practice, both can be regarded as definitions. 

Rather than start with the mathematical theory and try to interpret it, I adopt a specific, orthodox (q.v. Bub \cite{Bub}
) interpretation and seek to produce the mathematical structure appropriate to it. The result is essentially relativistic quantum mechanics, but with subtle and sometimes important differences. In contrast to standard quantum theory, the model is background-free in the sense that the physical metric is determined from measurement results, not from the properties of a prior space-time. space-time is thus seen as emergent rather than substantive. Hilbert space is finite dimensional, but a continuum of kets, $ \left|x\right\rangle$ for $x\in \mathbb{R}^{3}$, is defined using linear combinations of basis kets (similarly, 3D space does not depend on the coordinates used to describe it). A wave equation governing time evolution is not assumed as a postulate, but is established from probabilistic considerations. Momentum space is not assumed, since it is a part of the mathematical structure of Hilbert space.

This treatment is based on the observation that when there is no means, even in principle, to define the coordinates of a particle, quantum effects appear. The interpretation follows Dirac \cite{Dirac} 
and von Neumann \cite{von Neumann}
, has its origins in the Copenhagen interpretation as discussed by Heisenberg \cite{Heisenberg}
, and shares much with modern views such as Mermin \cite{Mermin}
, Adami and Cerf \cite{Adami}
, and Rovelli \cite{Rovelli1}
. As in the Copenhagen interpretation matter has an unknown but real behaviour which is not directly described by quantum mechanics. By giving a probability for each outcome, the ket describes not what is but our knowledge of what might happen in measurement; quantum theory is essentially a theory of probabilistic relationships between measurement results, not a model of physical processes between measurements.

The orthodox, or Dirac-von Neumann, interpretation should not be conflated with the Copenhagen interpretation, since Copenhagen invokes some notion of complementarity which is absent in Dirac-von Neumann. The interpretation here is orthodox, but goes further than both Dirac and von Neumann. For example, Dirac (quoted in section \ref{Relationism}) stated what cannot be said of quantum particles, but not what can be said; here a particle is defined as a physical entity in the absence of space-time background. Von Neumann described quantum logic as a language which tells us what can be discovered from measurement but he did not \textit{translate} the propositions of quantum logic into English. Similarly, Jauch \cite{Jauch} 
has described the propositional calculus as a foundation for quantum mechanics, but this is an abstract treatment inaccessible to many physicists. Here concrete propositions for quantum theory are abstracted directly from the formal statement of sentences in ordinary language.

The treatment given here neglects spin. The inclusion of spin raises additional issues concerning the interpretation of the projection postulate. By ignoring spin these issues do not arise. The present treatment is extended in \cite{Francis} where it is observed that spin is a required property of particles in relativistic quantum theory. Measurement issues concerning the reasons for the projection postulate as applied to spin depend upon the physical processes involved in measurement, and can only be resolved after considering quantum electrodynamics as a theory of interactions between particles.

\subsection{Relationism}
\label{Relationism}
Relationism is the principle that, since a measurement of distance is a comparison between the matter (and radiation) being measured and the matter (and radiation) it is measured against, only relative distances should appear at a fundamental level in physical theory. Although the mathematical formulation of physical law has depended on an assumption of space, or more recently space-time, imbued with mathematical properties, the Cartesian relationist view continues to hold intellectual appeal and, as described by Dieks \cite{Dieks}
, there is some reason, both within the foundations of quantum mechanics and in relativity, for thinking that the correct way to formulate physical theory would be to describe space-time as a collection of frame-dependent sets of potential measurement results, rather than as a background into which matter is placed in the manner of Newtonian space. In recent years relationism has been used by Smolin \cite{Smolin}
, Rovelli \cite{Rovelli2} 
and others as motivation for work on background-free theories such as spin networks, and has been suggested as basis for understanding quantum mechanics \cite{Rovelli1} 
and quantum gravity \cite{Poulin}
.

Relativity of motion is often stated, \textit{`you cannot say how something is moving unless you say how it is moving relative to other matter'}. The relationist view also requires relativity of position; \textit{`you cannot say where something is unless you say where it is relative to other matter'}. Relationism is also suggested by the orthodox, or Dirac-von Neumann, interpretation of quantum mechanics, that it only makes sense to talk of measured values when a measurement is actually done, or when the outcome of a measurement can be predicted with certainty. \textit{``In the general case we cannot speak of an observable having a value for a particular state, but we can \ldots\ speak of the probability of its having a specified value for the state, meaning the probability of this specified value being obtained when one makes a measurement of the observable''} --- Dirac \cite{Dirac}
. We may infer from Dirac's words that a precise value of position only exists when a measurement of position is performed or has a certain outcome, so that we can only talk about where a particle is found in measurement, not where it is in space.

\subsection{Quantum logic}
\label{Quantum Logic}
The central problem with relationism has been the difficulty in expressing it formally as axioms for use in mathematical argument. Whereas Newton was able to describe mechanics in three laws, the mathematical implications of relationism were, and have remained, obscure. Here Hilbert space is seen as a formal language which allows us to mathematically describe the behaviour of matter in a universe in which position exists only as a relative quantity (`behaviour' is intended to indicate change with respect to time, and should be understood without spacial connotations). Quantum logic (see e.g. Svozil \cite{Svozil}
) was introduced by Garrett Birkhoff and John von Neumann \cite{Birkhoff} 
and is sometimes described as applying counter-intuitive truth values to simple propositions. This paper will interpret kets as formal conditional clauses, rather than as propositions. The inner product combines clauses to generate formal propositions in the subjunctive mood, showing that the language is a consistent and intuitive extension of two-valued logic and classical probability theory and a natural formalisation of statements about measurements in the subjunctive mood. The principle of superposition is simply logical disjunction in formal language; there is no suggestion of an ontological quantity of magnitude $\left|\left\langle x|f\right\rangle\right|$ associated with a particular particle.

Classical probability theory can be used to describe physical situations where the outcome depends upon an unknown quantity, or random variable. In the absence of a random variable, the probability interpretation of quantum mechanics is physically unjustified. It will be seen in section \ref{Observable Quantities} that, for a normalised ket $ \lvert f \rangle $, the probability of a measurement result, $ k $, $ P(k|f) = \langle f|K|f\rangle$ can be understood as a classical probability function, where the random variable, $ K $, runs over the set of projection operators corresponding to the outcomes of the measurement. The physical interpretation is that each projection operator represents a set of unknown configurations of matter, namely that set of configurations leading to a given measurement result. Thus, it will be seen that Hilbert space is explained as the mathematical structure of a formal language, and that this language, when one learns to use it, can be used to describe the physical properties of a universe with no space-time background, and in which all properties are relationships between matter (or radiation) and other matter.

\subsection{Discreteness}
\label{Discreteness}
Since Newton, the continuum has been induced from the empirical accuracy of physical laws that use it for their expression. But, as Hume argued and Leibniz demonstrated, induction does not provide rigorous scientific proof because an indefinite number of laws can always be found to fit any finite body of data. In this paper the apparatus is not treated from a classical perspective, as in standard Copenhagen. We merely require that the result of measurement of position at given time is always three numbers, and use those numbers to label a condition found in matter. We assume measurement to a level of accuracy limited only by physical law and the ingenuity of the makers of the apparatus. In practice measurement results can always be expressed as terminating decimals, and we choose some bounding range and resolution at which to define a basis for a finite dimensional Hilbert space. We can, in principle, use resolutions greater than that of our current apparatus, but observation never permits us to say ``for all resolutions'' but only ``for resolutions up to the current limit of experimental accuracy'' (future technology may provide greater resolution, but in any future technology the resolution will still be finite, if only because we cannot write a number with infinite decimal places).

It is well understood that a discrete model cannot be manifestly covariant. Manifest covariance will not be applied since it is by definition the case that the apparatus is stationary with respect to the reference frame and affects the measurement result. By reference frame I do not mean coordinate system, but rather the chosen matter from which a coordinate system may be determined in practice by physical measurement, as in, e.g., ``the Earth frame'' or ``the frame of the fixed stars''. Since the reference frame is defined by the apparatus, it is meaningless to talk of rotations of the frame unless one is also rotating (or replacing) the apparatus. But in that case one is not rotating vector quantities, but rather redefining them in a new frame. Quantum covariance will also take into account that part of this effect is that the apparatus has a finite resolution, and will restore the principle that local laws of physics are the same in all reference frames. 

There are technical advantages in using finite dimensional Hilbert space in that stronger theorems are available and the order of taking limits can be tracked. In certain instances (loop integrals) the order of taking limits is critical as to whether the limit exists. It will be shown that discrete position functions for all coordinate systems are uniquely embedded in smooth wave functions. The continuum equations remove any dependency on a specific measurement apparatus and resolution because they contain embedded within them the solutions for all discrete coordinate systems possible in principal or in practice. Thus, in spite of discreteness, the theory is invariant under changes of basis. 

\section{Measurement}
\label{sec:Measurement}
\subsection{Reference matter}
\label{Reference Matter}
When a human observer seeks to quantify nature, he chooses some particular matter from which to define a reference frame or chooses certain matter from which he builds his experimental apparatus. He then observes a defined relationship between this specially, but arbitrarily, chosen reference matter and whatever matter is the subject of study. Here measurement is distinguished from a simple count of a number of objects, and is defined to mean a count of units of a measured quantity, where the definition of the unit of measurement invokes comparison between some aspect of the subject of measurement and a property of the reference matter used to define the unit of measurement. The division between reference matter and subject matter is present in all measurement and appears as the distinction between particle and apparatus in quantum mechanics, and in the definition of position relative to a reference frame in special relativity. 

Reference matter is to a large degree arbitrary, and is itself subject to measurement with respect to other matter. D'Inverno \cite{d'Inverno} 
defines a reference frame as a clock, a ruler, and coordinate axes, whereas Rindler \cite{Rindler} 
describes a reference frame as a ``conventional standard'' and discusses the attachment of a frame to definite matter, such as the Earth or the ``fixed'' stars, while Misner, Thorne and Wheeler \cite{Misner} 
define proper reference frame as a Minkowski coordinate system with a given clock at the origin. Whatever reference matter is used it includes some form of clock, axes, and some means of determining distance, such as a ruler or radar, and it may include any form of apparatus used for physical measurement. In all cases a property is measured relative to other, arbitrarily chosen matter, and the measurement determines a relationship between subject and reference matter, rather than an absolute property of the subject of measurement.
Inertial reference matter is assumed, where inertial is taken to mean that the effect on motion of contact interactions with other matter is negligible. Alternatively inertial coordinates may be calculated from the reference matter (e.g., a satellite spinning on its axis may be used to determine an inertial reference frame, although it is not itself inertial). This introduces complications in the description, but not complications of a fundamental nature. 

\subsection{Coordinates}
\label{Coordinates}
We are particularly interested in measurement of time and position. This is sufficient for the study of many (it has been said all) other physical quantities and we restrict our treatment to those physical quantities that can be reduced to a set of measurements of position, including measurements of position of particles other than the one under study, such as the position of a pointer. For example, a classical measurement of velocity may be reduced to a time trial over a measured distance, and a typical measurement of momentum of a particle involves plotting its path in a bubble chamber (or equivalent), being a set of positions over a time interval. 

Local distance measurements may be defined by the radar method. Any method of measuring coordinates may be used, calibrated to the radar method, so it is natural to use synchronous spherical coordinates with time as a parameter as in non-relativistic quantum mechanics. For convenience, Cartesian coordinates will be chosen. This simplifies certain formulae, but makes no fundamental difference to the treatment. Any apparatus has a finite resolution and the values written down are triplets of terminating decimals, which can be scaled to integers in units of some bounding resolution. Measured positions are always discrete values, determined by the range and resolution of a measurement apparatus. In practice it is simpler to use an equally spaced lattice, containing very large number $N$ positions given by decimals terminating at some value beyond the best available resolution of any existing apparatus. Margins of error and measurements at lower resolution can be represented using finite sets of such integers. In practice there is also a bound on magnitude. Without loss of generality the same bound, $\nu\in\mathbb{N}$, is used for each coordinate. Knowledge of the ket at any time is thus restricted to this set of triplets and the results of measurement of position are in a (subset of a) finite region, $\mathrm{D}\subset(\chi\mathbb{Z})^3$.\\
\\
\textbf{Postulate:} The \textbf{discrete space coordinate system} is $\mathrm{D}\equiv(-\chi \nu, \chi\nu]^3\subset(\chi\mathbb{Z})^3$ for some $\nu\in\mathbb{N}$, and for some lattice spacing $\chi\in\mathbb{Q}$ with $\chi>0$.\\

Let $\mathrm{T}\subset\chi\mathbb{Z}$ be a finite discrete time interval such that any particle under study will be measured in $\mathrm{D}$ for times $t\in\mathrm{T}$.\\
\\
\textbf{Postulate:} The \textbf{discrete space-time coordinate system} is $\mathrm{S}\equiv\mathrm{T}\otimes\mathrm{D}$ and is calibrated such that the speed of light is 1 radially to the origin.\\

The coordinate system is a lattice determined by practical considerations. The lattice should be understood as a product of the observers measurement apparatus, or reference frame, not as intrinsic to the objects being described within that frame.  Not every element of $\mathrm{D}$ need correspond to a possible measurement result, but $\mathrm{D}$ contains as elements or subsets the possible measurement results for a measurement of position with the chosen apparatus. There is no significance in the bound, $\nu$, of a given coordinate system. It is not intended to take either the limit $\nu\rightarrow\infty$ or $\chi\rightarrow 0$, but $\chi\nu$ is large enough to neglect the possibility of particles leaving $\mathrm{S}$. In practice this is always the case since data is discarded from any trial in which there is not both a well defined initial and final state; the probability amplitudes defined below relate to conditional probabilities such that both initial and final states are unambiguously determined (hence there is no detection loophole in Bell tests --- in the absence of unambiguous detection this model does not apply). 

\subsection{Particles}
\label{Particles}
It is sometimes assumed that a particle is localised in space, even if at unknown location. This is not the case here, since a value for the position observable is not assumed to exist between measurements.\\
\\
\textbf{Postulate:} A \textbf{particle} is any physical entity whose position can be measured at given time such that the result of such measurement is a value, $x\in\mathrm{D}$, or a neighbourhood $\left\{x\in\mathrm{D}\right\}$ of negligible size.\\
\\
\textbf{Postulate:} An \textbf{elementary particle} is one which cannot, even in principle, be subdivided into particles for which separate positions can be measured.\\
\\
It is not necessary to assume the existence of an elementary particle on metaphysical grounds. If there is such a thing as an elementary particle, then its theoretical properties may be determined, and if something in nature exhibits precisely those properties, then we will claim that it is an elementary particle. Quarks may be considered as elementary particles having separate positions in principle, but bound in practice.

\subsection{Many valued logic}
\label{Many Valued Logic}
Many valued logics \cite{Rescher} 
were introduced in the 1920s by Jan \L{}ukasiewicz \cite{Lukasiewicz} for dealing with the intuitive idea of degrees of certainty. Is has become widely recognised since Harold Jeffrey's publication of Theory of Probability \cite{Jeffreys} 
that probability theory is a many valued logic \cite{Jaynes}
. Another popular many valued logic, fuzzy logic, created by Lofti Zadeh \cite{Zadeh}
, has been used with considerable success in systems science for problems involving approximate reasoning based on imprecise information as is typically supplied by natural language.

Classical logic applies to sets of statements about the real world which are definitely true or definitely false. For example, when we make a statement,\\
\\
$\mathcal{P}(x) = \textit{The position of a particle is }x$,\\
\\
we tend to assume that it is definitely true or definitely false. Such statements are said to be \textbf{sharp} or \textbf{crisp}, meaning that they have truth values from the set $\{0,1\}$. If it is the case that $\mathcal{P}(x)$ is definitely either true or false then classical logic and classical mechanics apply. Similarly, probability theory gives \textbf{Bayesian truth values} from the continuous interval $[1, 0]$ to sentences in the future tense:\\
\\
$\mathcal{Q}(x) = \textit{When a measurement of position is done the result will be } x$.\\
\\
Similarly fuzzy logic assigns truth values on the interval $ [0, 1] $ to vague statements such as \textit{``he is a tall man''}. 

In quantum mechanics we deal with situations in which there has been no measurement and there is not going to be one. $\mathcal{P} (x)$ and $\mathcal{Q}(x)$ are not then legitimate propositions about physical reality. For example, we only get interference from Young's slits when there is no way to determine which slit the particle came through. In the absence of measurement we can consider propositions describing hypothetical measurement results, such as the set of propositions of the form:\\
\\
$\mathcal{R} (x) = \textit{If a measurement of position were done the result would be } x$.\\
\\
$\mathcal{R} (x)$ is intuitively sensible, even when no measurement is done, but cannot sensibly be given a crisp truth value. Its truth is distinguished from that of $\mathcal{Q} (x)$ because, when no measurements are to be done, we cannot sensibly discuss the potential frequency of individual measurement results. 

\subsection{Formal language}
\label{Formal Language}
In quantum theory we are not always going to do a measurement, but we want to talk about what would happen if we were to do a measurement, i.e. we need to be able to make statements about hypothetical measurement results. Hilbert space provides a way of discussing levels of truth for statements about hypothetical measurement, like $\mathcal{R}(x)$, in the subjunctive mood. Statements in the subjunctive consist of two clauses, the conditional clause \textit{``If a measurement of position were done, \ldots''}, and the consequent clause \textit{``\ldots, then the result would be $x$''}. The conditional clause will contain whatever information is known from prior measurement. We therefore discuss two measurements, the first to determine the condition and the second to determine the outcome, or consequence. We represent the results of these measurements symbolically. The conditional clause, referring to the first measurement, is represented by a ket. It is described as a formal conditional clause to indicate that only clauses formally described in the rules are allowed in formal language. \textbf{Basic conditional clauses}, on which the language is built, refer directly to individual measurements of position:\\
\\
\textbf{RULE I.} For $x\in\mathrm{D}$, $|x\rangle$  is the \textbf{formal conditional clause} \textit{``If measured position at time $t$ were $x$, \ldots''}.\\
\\
An actual position found by a real apparatus is described by a set of points in the lattice. To describe this we need to extend the language, by introducing an operator corresponding to \textsc{or}, represented by the symbol $+$. To express the idea that one possibility is more likely than the other we introduce a weighting. Thus, if the magnitude of $a$ is greater than that of $b$, then $a|g\rangle + b|f\rangle$ will mean \textit{``if measured position were either $x$ or $y$, but more likely $x$, \ldots''}. We also want to be able to express many possibilities, \textit{``If the particle were found at $x$ or $y$ or $z$ or \ldots''}. This is done recursively in rule II:\\
\\
\textbf{RULE II.} If $|g\rangle$ and $|f\rangle$ are formal conditional clauses, and $a$ and $b$ are complex numbers, then $a|g\rangle + b|f\rangle$ is a formal conditional clause.\\
\\
The set of formal conditional clauses, or kets, now has the mathematical structure of an $N$-dimensional vector space, $\mathbb{H}^1(t)$, where $N=8\nu^3$. The elements of $\mathbb{H}^1(t)$ are formal conditional clauses concerning the measurement of position of a single particle at time $t$. Basic conditional clauses, $|x\rangle$, are a basis for $\mathbb{H}^1(t)$. Kets are not strictly states of a particle, but formal conditional clauses describing hypothetical measurement results. They will be referred to as ``states'', in keeping with common practice when no confusion arises. The use of a vector space over the complex numbers introduces a degree of freedom which will be used in the description of the evolution of kets.

To complete a formal sentence we need to put a formal conditional clause together with a formal consequent clause. Consequent causes refer to a second measurement, at the same time as the first measurement. To make statements about real measurement results we will also need to know how kets evolve in time, but in the first instance the discussion is restricted to hypothetical measurements at time $t$. There is no fundamental difference between one measurement and another, so the grammatical structure, \textit{weighted disjunction}, described in rule II, applies equally well to consequent clauses. These also form an $N$-dimensional vector space, defined from a basis of consequent clauses in one-one correspondence with the basic conditional clauses, or kets, described by rule I. Consequent clauses are represented symbolically by bras:\\
\\
\textbf{RULE III.} $\langle x|$ is the \textbf{formal consequent clause} \textit{``\ldots, then, in a second measurement at time $t$, measured position would be $x$''}.\\
\\
We put the two clauses together, to make a braket, representing a statement about measurement at a given time:\\
\\
\textbf{RULE IV.} $\langle x|y\rangle$ is the statement \textit{``If measured position at time $t$ were $y$, then, in a second measurement at time $t$, measured position would be $x$''}.\\
\\
From observation we know that, if, at some particular time, a particle is measured at position $x$, then its position is definitely $x$ and it cannot be measured separately at some other position $y$ at the same time. The statement $\langle x|y\rangle$ is strictly true or false, depending on whether or not $x = y$.\\
\\ 
\textbf{Postulate:} The \textbf{truth value} of $\langle x|y\rangle$ is given by a Kronecker delta, $\langle x|y\rangle = \delta_{xy}$.\\
\\
With linearity and complex conjugation, this defines an inner product between any two kets, $|f\rangle,|g\rangle \in \mathbb{H}^1(t)$. Thus, $\mathbb{H}^1(t)$ is a Hilbert space, the basic conditional clauses of rule I are an orthonormal basis, and the space of bras is the dual space. In effect propositions in the subjunctive have complex truth values. This extends the usual definition of a many valued logic in which truth values are real. However, a truth value for a statement about hypothetical measurement has no direct meaning in the real world, but is defined to be whatever we choose it to be. Whether or not we describe complex values of the inner product as ``truth values'' is inconsequential.  \\
\\
\textbf{Definition:} The \textbf{position function} of the ket $|f\rangle \in \mathbb{H}^1(t)$ is the mapping, $\mathrm{D}\rightarrow\mathbb{C}$, $\forall x \in \mathrm{D}, x \rightarrow \langle x | f \rangle $.\\
\\
Later the position function will be identified with the restriction of the wave function to $\mathrm{D}$. It is here termed ``position function'' because it is discrete and because a wave equation is not assumed. 

In this formal language, relative magnitudes are important in weighted logical \textsc{or}, but absolute magnitude has no meaning. It is easy in common language to construct phrases containing redundant words. \textit{``The black piece of coal''} is not the same phrase as \textit{``the piece of coal''}, but both have the same meaning. Similarly, for any complex number $a$, the clause $|f\rangle$ means exactly the same thing as $a|f\rangle$. When not part of a larger construction containing $+$, $a$ has the role of a redundant word.

The resolution of unity is found by expanding a ket in a normalised basis
\begin{equation}\label{Eq3}
\ | f \rangle = \sum_{x \in \mathrm{D}} | x \rangle \langle x | f \rangle.
\end{equation}
Hence
\begin{equation}\label{Eq4}
\ 1 = \sum_{x \in \mathrm{D}} | x \rangle \langle x | .
\end{equation}
The inner product is strictly a finite sum with $N$ terms, where $N = 8\nu^3$ is large. The formal limit $N\rightarrow \infty$, $\chi\rightarrow 0$ is only to be taken at the final stage of calculation. With this in mind, it is convenient to normalize basis kets,
\begin{equation}\label{Eq5}
\forall x,y \in \mathrm{D},  \langle x | y \rangle = \chi^{-3} \delta_{xy}.
\end{equation}
With this normalisation, the resolution of unity takes the form:
\begin{equation}\label{Eq6}
\ 1 = \chi^3 \sum_{x \in \mathrm{D}} | x \rangle \langle x | .
\end{equation}

\subsection{Multiparticle kets}
\label{Multiparticle kets}
\begin{flushleft}
\textbf{RULE Va.} $|\rangle$ is the formal conditional clause, \textit{``If the first measurement at time $t$ were to find no particle, \ldots''}.\\
\end{flushleft}\textbf{RULE Vb.} $\langle |$ is the formal consequential clause, \textit{``\ldots, then a second measurement at time $t$ would find no particle''}.\\
\\
\textbf{Definition:} Let $\mathbb{H}^0$ be the space spanned by $|\rangle$.\\
\\
Because multiplication by scalars only has meaning in association with the weighting in \textsc{or}, there is no difference in meaning between member clauses, $a|\rangle$, of $\mathbb{H}^0$.\\
\\
\textbf{Postulate:} The space of kets for $n$ particles of the same type is given by the $n^{th}$ tensor power $\mathbb{H}^n \equiv (\mathbb{H}^1)^{\otimes n} \equiv \underbrace{\mathbb{H}^1\otimes\cdots\otimes \mathbb{H}^1}_{n}$\\
\\
\textbf{RULE VIa.} $|x_1\rangle |x_2\rangle \ldots |x_n\rangle$ is the formal conditional clause, \textit{``If, for each of $n$ particles, the measured position at time $t$ of the $i^{th}$ particle were $x_i$, \ldots''}.\\
\\
\textbf{RULE VIb.} $\langle x_1|\langle x_2|\ldots\langle x_1|$ is the formal consequential clause, \textit{``\ldots, then, for each of $n$ particles in a second measurement at time $t$, the measured position of the $i^{th}$ particle would be $x_i$''}.\\
\\
\textbf{Postulate:} The space of any number of particles of the same type, $\gamma$, is $\mathbb{H}_\gamma \equiv \bigoplus\limits_{n} \mathbb{H}^n$.\\
\\
The direct sum allows statements about an uncertain number of particles, using weighted logical \textsc{or}, \textit{``If, for each of $n$ or $m$ particles, but more likely $n$ than $m$, \ldots''}, etc. Since an $n$ particle ket cannot be an $m$ particle ket, the braket between kets of different numbers of particles is zero. For $|f\rangle=|f_1\rangle\ldots|f_n\rangle \in \mathbb{H}^n$, $|g\rangle=|g_1\rangle\ldots|g_n\rangle \in \mathbb{H}^n$,
\begin{equation}\label{Eq6.5}
\langle f | g \rangle = \prod_{i=1}^{n} \langle f_i | g_i \rangle ,
\end{equation}
as is required for independent particles by the probability interpretation (section \ref{Probability Interpretation}). \\
\\
\textbf{Postulate:} The space of particles is $\mathbb{H} \equiv \bigoplus\limits_{\gamma} \mathbb{H}_{\gamma}$.\\
\\
\textbf{RULE VIIa.} $|x_1;x_2;\ldots;x_n \rangle$ is the formal conditional clause \textit{``If, for $n$ identical particles, measured positions at time $t$ were $x_1$, $x_2$, \ldots, $x_n$''}.\\
\\
\textbf{RULE VIIb.} $\langle x_1;x_2;\ldots;x_n |$ is the formal consequential clause \textit{``then, for $n$ identical particles, measured positions at time $t$ would be $x_1$, $x_2$, \ldots, $x_n$''}.\\
\\
\textbf{Postulate:} Since switching identical particles makes no difference to the physical situation, multiparticle space is \textbf{Fock space}, $\mathbb{F} \equiv \bigoplus\limits_{n} S\mathbb{H}^n$ where $S$ means that groups of tensor indices referring to the same type of particle are symmetrised for Bosons and antisymmetrised for Fermions. 

\section{Momentum space}
\label{Momentum Space}
\subsection{Formal definition}
\begin{flushleft}
\label{Formal definition}
\textbf{Definition:} For a 3-vector, $p$, at the origin, define the \textbf{momentum ket}, $|p\rangle$, as a sum of position kets:
\begin{equation}\label{Eq3.1.1}
| p \rangle = \left(\tfrac{1}{2\pi}\right)^{3/2}\chi^3\sum_{x\in\mathrm{D}}e^{ix\cdot p}|x\rangle,
\end{equation}
\end{flushleft}
where the dot product uses the Euclidean metric. The Euclidean metric in \eqref{Eq3.1.1} has no direct bearing on a physical metric, and merely defines momentum kets as linear combinations of basic conditional clauses. The inner product with $|x\rangle$ defines a \textbf{plane wave},
\begin{equation}\label{Eq3.1.2}
\langle x | p \rangle = \left(\tfrac{1}{2\pi}\right)^{3/2}e^{ix\cdot p}.
\end{equation}
\textbf{Definition:} $|p\rangle$ is a \textbf{plane wave ket} with \textbf{momentum} $p$. \\
\\
This is the fundamental definition of 3-momentum in this approach. It is justified because it is found in qed that $p$ is a conserved quantity which corresponds precisely to the classical notion of momentum \cite{Francis}. In this paper only Newton's first law will be shown.\\
\\
\textbf{Definition:} \textbf{Continuum momentum space} is the 3-torus, $\mathrm{M} \equiv (-\tfrac{\pi}{\chi},\tfrac{\pi}{\chi}]^3 \subset \mathbb{R}^3$.\\
\\
There are momentum kets $|p\rangle$ in $\mathbb{H}^1$ for continuum values of $p\in\mathrm{M}$ (since they're just linear combinations of basis kets $|x\rangle$), but a discrete subset of momentum kets,
\begin{equation}\label{Eq9}
\left\{|p\rangle, p\in \mathrm{M}_{\mathrm{D}} = \mathrm{M} \cap (\chi_p \mathbb{Z})^3\right\},
\end{equation}
is a basis for $\mathbb{H}^1$, where lattice spacing for $\mathrm{M_D}$ is given by $\chi_p = \pi / (\chi\nu)$. Using discrete transforms, Fourier inversion is exact. The resolution of unity in momentum space is
\begin{equation}\label{Eq10}
\chi_p^3 \sum_{p\in \mathrm{M_D}} |p\rangle \langle p | = 1.
\end{equation}\\
\\
\textbf{Definition:} For $|f\rangle \in \mathbb{H}^1(t)$, determined by measurement at time $x^0 = t$ using discrete coordinates, $\mathrm{D}$, the \textbf{momentum space wave function} $F:\mathrm{M} \rightarrow \mathbb{C}$ is $p \rightarrow F(p) = \langle p | f \rangle$.\\
\\
In particular, for the position ket $|z\rangle$, the momentum space wave function is, for $p \in \mathrm{M}$,
\begin{equation}\label{Eq12}
p \rightarrow \langle p | z \rangle = \left(\tfrac{1}{2\pi}\right)^{3/2}e^{-iz\cdot p}.
\end{equation}
It is straightforward to show that, for $x,y \in \mathrm{D}$,
\begin{equation}\label{Eq13}
\int_{\mathrm{M}}d^3p \,\langle x|p \rangle\langle p|y \rangle = \left(\tfrac{1}{2\pi}\right)^{3}\int_{\mathrm{M}}d^3p\, e^{-iy\cdot p}e^{ix\cdot p} = \chi^{-3} \delta_{xy} = \langle x|y \rangle.
\end{equation}
Thus, Fourier inversion holds using the integral on momentum space; for any $|f\rangle \in \mathbb{H}^1(t)$,
\begin{equation}\label{Eq14}
\int_{\mathrm{M}}d^3p \,\langle x|p \rangle\langle p|f \rangle =  \int_{\mathrm{M}}d^3p\, \chi^3 \sum_{y \in \mathrm{D}} \langle x|p \rangle\langle p|y \rangle \langle y| f \rangle = \langle x| f \rangle.
\end{equation}
We can thus identify the sum over discrete momenta with an integral over $\mathrm{M}$,
\begin{equation}\label{Eq15}
1 \equiv \chi_p^3\sum_{p \in \mathrm{M_D}} |p \rangle\langle p| \equiv \int_{\mathrm{M}} d^3p \,|p \rangle\langle p|.
\end{equation}
Then for any $|f\rangle \in \mathbb{H}^1(t)$, $q \in \mathrm{M}$
\begin{equation}\label{Eq16}
\langle q|f\rangle \equiv \chi_p^3\sum_{p \in \mathrm{M_D}} \langle q|p \rangle\langle p|f\rangle \equiv \int_{\mathrm{M}} d^3p\, \langle q|p \rangle\langle p|f\rangle.
\end{equation}
Thus, for any $p,q \in \mathrm{M}$, $\langle q|p\rangle = \delta(p-q)$. It is perhaps unexpected that the Dirac delta function on the test space of momentum space wave functions has an exact representation as a smooth function, 
\begin{equation}\label{Eq16.3}
\delta(p-q) \equiv \left(\tfrac{1}{2\pi}\right)^{3} \chi^3 \sum_{x \in \mathrm{D}}e^{ix\cdot (p-q)}.
\end{equation}

\subsection{Smooth representation}
\label{sec3.2}
\begin{flushleft}
\textbf{Definition:} $\mathrm{D}$ is embedded into the \textbf{continuum coordinate system}, $\mathrm{C}$,
\begin{equation}\label{Eq:3.2.1}
\mathrm{D} \subset \mathrm{C} \equiv (-\chi\nu, \chi\nu]^3 \subset \mathbb{R}^3.
\end{equation}
\end{flushleft}
\textbf{Definition:} For any $x \in \mathrm{C}$ we may define the \textbf{position ket} 
\begin{equation}\label{Eq:3.2.2}
|x\rangle = \chi_p^3 \sum_{p \in \mathrm{M_D}} |p\rangle \langle p | x \rangle = \int_{\mathrm{M}} d^3p\, |p \rangle \langle p | x \rangle.
\end{equation}
\textbf{Definition:} The \textbf{wave function} for $|f(t)\rangle \in \mathbb{H}^1(t)$ is $f(t): \mathrm{C} \rightarrow \mathbb{C}$ with
\begin{equation}\label{Eq:3.2.3}
x \rightarrow f(t,x) = \langle x|f(t) \rangle = \chi^3 \sum_{z \in \mathrm{D}} \langle x|z \rangle \langle z | f(t) \rangle.
\end{equation}
Expanding the wave function in momentum space gives, for $x\in \mathrm{C}$, 
\begin{equation}\label{Eq:3.2.4}
f(x)=\langle z|f\rangle = \int_{\mathrm{M}}d^3p\, \langle x|p \rangle \langle p | f \rangle = \left(\tfrac{1}{2\pi}\right)^{3/2} \int_{\mathrm{M}}d^3p\, e^{ix\cdot p}\langle p | f \rangle ).
\end{equation}
Wave functions are differentiable. The wave function for $|z\rangle$, $z\in \mathrm{C}$, is, for $x\in \mathrm{C}$,
\begin{equation}\label{Eq:3.2.5}
x \rightarrow f_z(x) = \int_{\mathrm{M}}d^3p\, \langle x|p \rangle \langle p | z \rangle = \left(\tfrac{1}{2\pi}\right)^3 \int_{\mathrm{M}}d^3p\, e^{i(x-z)\cdot p}
\end{equation}
It is easily verified that for $x,z \in \mathrm{D}$ $f_z(x)=\chi^{-3}\delta_{xz} = \langle x | z \rangle$. So, the position function is the restriction of the wave function to $\mathrm{D}$, and, for $z \in \mathrm{D}$, there is a one-one correspondence between the wave functions, $f_z(x)$, and basis kets, $|z\rangle$, such that smooth wave functions are a representation of a finite dimensional Hilbert space. For $p, q \in \mathrm{M}$
\begin{equation}\label{Eq:3.2.6}
\int_{\mathrm{C}} d^3x \langle p|x \rangle \langle x | q \rangle = \left(\tfrac{1}{2\pi}\right)^{3}\int_{\mathrm{C}} d^3x\, e^{-ix\cdot (p-q)} = \chi_p^{-3} \delta_{pq} = \langle p|q \rangle.
\end{equation}
So, by linearity, we can identify the sum over discrete coordinates with an integral. The identity operator $1:\mathbb{H}^1 \rightarrow \mathbb{H}^1$ can be written
\begin{equation}\label{Eq:3.2.7}
1 \equiv \chi^3 \sum_{x \in \mathrm{D}} |x \rangle \langle x | \equiv \int_{\mathrm{C}} d^3x\, |x \rangle \langle x |.
\end{equation}
Then for any $|f\rangle \in \mathbb{H}^1$, $y \in \mathrm{C}$
\begin{equation}\label{Eq:3.2.8}
\langle y|f \rangle = \chi^3 \sum_{x \in \mathrm{D}} \langle y |x \rangle \langle x |f\rangle = \int_{\mathrm{C}} d^3x\, \langle y |x \rangle \langle x |f \rangle.
\end{equation}
and for any $x,y \in \mathrm{C}$ $\langle x|y \rangle = \delta(x-y)$ where the Dirac delta is a smooth function:
\begin{equation}\label{Eq:3.2.9}
\delta(x-y) \equiv (\tfrac{\chi_{p}}{2\pi})^3 \sum_{p \in \mathrm{M_D}} e^{i(x-y)\cdot p} \equiv \int_{\mathrm{M}} d^3p \, e^{i(x-y)\cdot p} .
\end{equation}

\subsection{Bounds}\label{sec3.3}
Since coordinate space is discrete, momentum space is the 3-torus $\mathrm{M}$, which is not covariant. The theory would break down if physical momentum could exceed $p_{\mathrm{max}}=\pi/\chi$, where $\chi$ is the lower bound of small lattice spacing, not the spacing appropriate to a given apparatus. In conventional units the components of momentum have a theoretical bound $p_{\mathrm{max}}=\pi\hbar c/\chi$. If Planck length is the smallest unit inherent in nature, the theoretical bound on the energy of an electron is $3.8 \times 10^{28}$eV, well beyond any reasonable level. Thus, in practice, physical momentum does not approach the bound and there is not an issue. 

In fact, there is a much lower bound on energy-momentum since an interaction between a sufficiently high energy electron and any electromagnetic field leads to pair creation (the Greisen-Zatsepin-Kuz'min limit on the energy of cosmic rays is $5 \times 10^{19}$eV \cite{Greisen}
\cite{Zatsepin}
). It follows from conservation of energy that the total energy of a system is bounded provided that energy has been bounded at some time in the past. This is true whenever an energy value is known since a measurement of energy creates an eigenket with a definite value of energy. Then momentum is also bounded, by the mass shell condition. The probability of finding a momentum above the bound is zero, and we assume that, for physically realizable states, $\langle p|f \rangle$ vanishes above the bound on each component of momentum. The bound depends on the system under consideration, but without needing to specify a least bound, we may reasonably assume that momentum is always much less than $\pi /(4\chi)$.

A theoretical bound on momentum might introduce a problem of principle for Lorentz transformation. If a high energy electron were boosted beyond the bound it might appear after the boost with a low energy, or with opposite direction of momentum. However, realistic Lorentz transformation means that macroscopic matter (i.e. the reference frame) is physically boosted by the amount of the transformation. In practice, Lorentz transformation cannot boost momentum beyond the level for which it is consistently defined.

The non-physical periodic property of $\langle p|f \rangle$  can removed by the substitution $\Theta_{\mathrm{M}}(p)\langle p|f \rangle \rightarrow \langle p|f \rangle$, where $\Theta_{\mathrm{M}}(p)=1$ if $p\in\mathrm{M}$ and $\Theta_{\mathrm{M}}(p)=0$ otherwise. With the replacement of the Euclidean dot product with Minkowski dot product (which takes place naturally in the solution of the Schr\"{o}dinger equation, section \ref{sec5.1}), the expansion of the wave function in momentum space \eqref{Eq:3.2.4} is identical to the standard form in relativistic quantum mechanics, up to normalisation, and can be put into a manifestly covariant form:
\begin{equation}\label{Eq:3.3.1}
\begin{split}
f(x) &= \left(\tfrac{1}{2\pi}\right)^{3/2} \int_{\mathbb{R}^3}d^3p\, \langle p|f \rangle e^{-ix\cdot p}\\
&= \left(\tfrac{1}{2\pi}\right)^{3/2} \int_{\mathbb{R}^3}\frac{d^3p}{2p^0} F(p) e^{-ix\cdot p}\quad  \text{where}\quad  F(p)=2p^0\langle p|f \rangle \\
&= \left(\tfrac{1}{2\pi}\right)^{3/2} \int_{\mathbb{R}^4}d^4p F(p) e^{-ix\cdot p}\delta(p^2-m^2).
\end{split}
\end{equation}

\section{Observable quantities}
\label{Observable Quantities}

\subsection{Probability interpretation}
\label{Probability Interpretation}

To make the formal language precise, we must assign numerical values to the complex numbers introduced in rule II, i.e. we must determine magnitude and phase. Phase contains information on the evolution of kets, and will be considered later. Magnitude will be determined from probability. It only makes sense to talk about probability when we are actually going to do a measurement. When we are actually going to do the measurement, a statement about hypothetical measurement, in the subjunctive mood, automatically becomes a statement about real measurement, in the future tense. This being the case, truth values for hypothetical results must be replaced by truth values for future events, i.e. probabilities, when experiments are actually done.

In a typical measurement in quantum mechanics we study a particle in near isolation. The suggestion is that there are too few ontological relationships to create the property of position and that measurement introduces interactions which generate position. In this case, prior to measurement, position does not exist and the state of the system is not labelled by a position ket. Instead, Hilbert space is used to provide a label containing information about the about the probability of what would happen in measurement. To associate a ket, $|f\rangle$, with a particular physical state it is necessary and sufficient to specify the magnitude and phase of $\langle x|f\rangle$ from empirical data. If we set up many repetitions of a system described by the initial measurement results, $f$, and record the frequency of each result, $x$, then for a large number of repetitions the relative frequency of $x$ tends to the probability, $P(x|f)$, of finding the particle at $x$. Thus, in the first instance, amplitudes of the components $\langle x|f\rangle$ are determined from the probabilities of measurement results, not the other way about. In practice they are determined from the results of previous measurements for which the results are known, together with the Schr\"{o}dinger equation (section \ref{sec5.1}).\\
\\
\textbf{Postulate:} For the ket $|f\rangle \in \mathbb{H}^1(t)$, the magnitudes of the coefficients, $\langle x|f\rangle$ are defined such that
\begin{equation}\label{Eq4.1.1}
\frac{|\langle x|f\rangle |^2}{\langle f|f\rangle} = P(x|f).
\end{equation}\\
\\
\textbf{Definition:} If $\langle f|f\rangle =1$ then $ |f\rangle $ is said to be \textbf{normalised}.

\subsection{Measurement}
\label{sec4.2}

Since only a general principle has been used that it is possible to measure position, it is necessary to discuss other observables. The question as to what other observables exist cannot be discussed until after a treatment of interactions between particles which goes beyond the scope of this paper. It is assumed that all observables are a product of physical laws arising from particle interactions. A full analysis of a given measurement would require that the measurement apparatus as well as the system being measured be treated as a multiparticle system in Fock space, in which time evolution for the interacting theory is known. Here general considerations are discussed on the assumption that interactions will be described by linear maps on Fock space and that measurement is always a physical process describable in principle as a combination of interaction operators (for qed this means that all observables depend only on the electric current operator and the photon field operator \cite{Francis}). A complete resolution of the measurement problem would demonstrate the projection postulate for any given apparatus and has not been given. The argument given below makes the projection postulate reasonable by reducing all measurement to measurement of position. The view is that if we find a physical process satisfying the projection postulate then we may say it defines an observable quantity.

Measurement has two effects on the state of a particle, altering it due to the interaction of the apparatus with the particle, and also changing the information we have about the state. New information causes a change of state even in the absence of physical change because the state is just a label for available information. Then the collapse of the wave function is in part the effect of the apparatus on the particle, and in part the effect on conditional probability when the condition becomes known. This inverts the measurement problem; collapse represents a change in information due to a new measurement but Schr\"{o}dinger's equation requires explanation --- interference patterns are real. The requirement for a wave equation will be found in section \ref{sec5.1}. 

Classical probability theory describes situations in which every parameter exists, but some are not known. Probabilistic results come from different values taken by unknown parameters. We have a similar situation here, but now the unknowns are not describable as parameters. We assume no relationships between particles bar those generated by physical interaction. An experiment is described as a large configuration of particles incorporating the measuring apparatus as well as the process being measured. The configuration has been partially determined by setting up the experimental apparatus, reducing the possibilities to those with definite outcomes to the measurement. It is impossible, even in principle, to determine every detail of the configuration since the determination of each detail requires measurement, which in turn requires a larger apparatus containing new unknowns in the configuration of particles. Thus there is always a lack of determination of initial conditions leading to randomness in the outcome, whether or not there is a fundamental indeterminism in nature. 

When we do a measurement, $K$, we get a definite result, a terminating decimal or $n$-tuple of terminating decimals read off the measurement apparatus. Let the possible results be $k_i \in \mathbb{Q}^n$ for $i=1,\ldots,m$. We assume that the dimension of $\mathbb{H}^1$ is greater than $m$; this must be so if all measurements are reducible to measurements of position, and can be ensured by the choice of a lattice finer than the resolution of measurement. Each physical state is associated with a ket, labelled by the measurement result, so that if the measured result is $k_i$ then the ket is $|k_i \rangle$. The empirical determination of $|k_i \rangle$ as a member of $\mathbb{H}^1$ requires that we draw from experimental data the value of the inner product $\langle k_i | f \rangle$ for an arbitrary ket, $|f \rangle$. Without loss of generality $|k_i \rangle$ and $|f \rangle$ are normalised. By assumption, measurement of $K$ is reducible to a set of measurements of position, so that each $k_i$ is in one to one correspondence with the positions $y_i$ of one or more particles used for the measurement (e.g. $y_i$ may be the positions of one or more pointers). Then, 
\begin{equation}\label{Eq4.2.1}
|\langle k_i|f\rangle |^2 = |\langle y_i|f\rangle |^2 = P(y_i|f) = P(k_i|f)
\end{equation}
is the probability that a measurement of $K$ has result $k_i$, given the initial ket $|f\rangle \in \mathbb{H}^1$. It follows from $\langle x|y \rangle = \delta_{xy}$ that $\langle k_i|k_j \rangle = \delta_{ij} = \langle y_i|y_j \rangle$. So, if the result is $k_i$ it is definitely $k_i$ and cannot at the same time be $k_j$ with $i\neq j$.

Measurement with result, $k_i$, implies a physical action on a system and is represented by the action of an operator, $K_i$, on Hilbert space. If a quantity is measurable we require that there is an element of physical reality associated with its measurement, by which we mean that the configuration of particles necessarily becomes such that the quantity has a well defined value. In practice this means that, in the limit in which the time between two measurements goes to zero, a second measurement of the quantity necessarily gives the same result as the first. It follows that $K_i$ is a projection operator (the projection postulate), 
\begin{equation}\label{Eq4.2.2}
K_i = | k_i \rangle \langle k_i |
\end{equation}
The projection postulate is too restrictive to describe all numerical quantities used in the classical description of nature, and will be relaxed after a discussion of expectations (section \ref{sec4.5}).

\subsection{Observable operators}
\label{sec4.3}

The expectation of the result from a measurement of $K$, given the initial normalised state, $|f\rangle \in \mathbb{H}^1$, is
\begin{equation}\label{Eq4.3.1}
\langle K \rangle \equiv \sum_i k_i P(k_i| f) = \sum_i \langle f|k_i \rangle k_i \langle k_i | f \rangle = \langle f|K|f \rangle
\end{equation}
\\
\textbf{Postulate:} The Hermitian operator, $K = \sum_i |k_i \rangle k_i \langle k_i | $, is called an \textbf{observable}. $k_i$ is the \textbf{value} of $K$ in the state $|k_i\rangle$.\\

Using \eqref{Eq4.2.1} the probability that operators describing the interactions comprising the measurement of $K$ combine to give the result $K_i$ is
\begin{equation}\label{Eq4.3.2}
P(k_i|f)=|\langle k_i | f \rangle |^2 = \langle f | k_i \rangle \langle k_i | f \rangle = \langle f | K | f \rangle .
\end{equation}
Then $P(k_i|f)$ can be understood as a classical probability function, where the random variable runs over the set of projection operators, $K_i$, corresponding to the outcomes of the measurement. The physical interpretation is that each $K_i$ represents a set of unknown configurations of particle interactions in measurement, namely that set of configurations leading to the result $k_i$. 

\subsection{The canonical commutation relation}\label{sec4.4}
\begin{flushleft}
\textbf{Definition:} The \textbf{momentum operator}, $P^a=-i\partial^a:\mathbb{H}^1 \rightarrow \mathbb{H}^1$, is, for $a=1,2,3,$  
\begin{equation}\label{Eq:4.4.1}
P^a:|f\rangle\rightarrow -\int_{\mathrm{C}}d^3x\,|x\rangle i\partial^a \langle x|f \rangle
\end{equation}
\end{flushleft}
Clearly $P^a$ is Hermitian and
\begin{equation}\label{Eq:4.4.2}
P^a|f\rangle = -\int_{\mathrm{C}}d^3x\,|x\rangle i\partial^a \chi_p^3 \sum_{p \in \mathrm{M_D}} \langle x|p \rangle \langle p|f \rangle = \chi_p^3 \sum_{p \in \mathrm{M_D}} |p \rangle p^a \langle p|f \rangle .
\end{equation}Similarly,
\begin{equation}\label{Eq:4.4.3}
P^a|f\rangle = \int_{\mathrm{M}}d^3p\, |p \rangle p^a \langle p|f \rangle.
\end{equation}
\textbf{Definition:} The \textbf{position operator}, $X^a:\mathbb{H}^1 \rightarrow \mathbb{H}^1$, is, for $a=1,2,3$
\begin{equation}\label{Eq:4.4.4}
X^a |f\rangle = \chi^3 \sum_{x \in \mathrm{D}} |x \rangle x^a \langle x|f \rangle
\end{equation}

From the property that the trace of a commutator in finite dimensional Hilbert space vanishes, $ \mathrm{Tr}([X^a,P^b]) = 0$, it follows that $[X^a,P^b]\neq i \delta_{ab}$, and the canonical commutation relation does not hold. If we formally define $\tilde{X}$ by
\begin{equation}\label{Eq4.4.4}
\tilde{X}^a|f\rangle = \int_{\mathrm{C}}d^3x \,|x \rangle x^a \langle x|f \rangle.
\end{equation}
Then,
\begin{equation}\label{Eq4.4.5}
P^b\tilde{X}^a|f\rangle = \int_{\mathrm{C}}d^3x \,|x \rangle i\delta_{ab} \langle x|f \rangle - \int_{\mathrm{C}}d^3x\, |x \rangle x^ai\partial^b \langle x|f \rangle = -i\delta_{ab} - \tilde{X}^aP^b|f\rangle.
\end{equation}
So,
\begin{equation}\label{Eq4.4.6}
[\tilde{X}^a,P^b]= i \delta_{ab}.
\end{equation}
and we conclude that $X^a \neq \tilde{X}^a$ and that $\tilde{X}^a|f\rangle \notin \mathbb{H}^1$. 

\subsection{Classical correspondence}
\label{sec4.5}

In the classical correspondence we study the behaviour of systems containing a large number, $N$, of quantum motions (this is sometimes called the thermodynamic limit). A classical property is the expectation, \eqref{Eq4.3.1}, of the corresponding observable in the limit $N \rightarrow \infty$ (not $\hbar \rightarrow 0$ as sometimes stated; Planck's constant is simply a change of scale from natural to conventional units and it would be meaningless to let it go to zero). For example, the centre of gravity of a macroscopic body is a weighted average of the positions of the elementary particles which constitute it. Schr\"{o}dinger's cat is definitely either alive or dead because, consisting as it does of a large number of elementary particles, its properties are expectations obeying classical laws derived from \eqref{Eq4.3.1}, but the ket simply encodes probability and the cat may be described as a superposition until the box is opened. A precise treatment of the time evolution of classical quantities requires the prior development of an interacting theory which will be the subject of a subsequent paper. It will be shown there that determinate laws obtain for classical quantities. In this paper we will simply assume determinate laws for expectations in the large number limit.\\
\\
\textbf{Postulate: }A \textbf{measurement} of a physical quantity is any physical process such that a determination of the quantity is possible in principle. \\

In keeping with the considerations of section \ref{sec4.2}, we assume that the existence of a value for an observable quantity depends only on the configuration of matter. If a configuration of matter corresponds to an eigenket of an observable operator then the value of that observable exists independently of observation and is given by the corresponding eigenvalue. In classical physics there is sufficient information to determine the motion at each instant between the initial and final ket, up to experimental accuracy. Intermediate kets are similarly determinate and may be calculated in principal by the processing of data already gathered, or which could be gathered without physically affecting the measurement. So in classical physics intermediate states may be regarded as measured states, and we may say that they are \textbf{effectively measured}, meaning that measurements on them have certain outcome. 

The projection postulate is required if the results of measurement are to be used to name states in Hilbert space, but classical quantities can also be defined from Hermitian operators when this is not the case. To say that a Hermitian operator has a well defined value in a given state, a measurement should necessarily yield that value as the expectation of the operator\\
\\
\textbf{Postulate:} For kets consisting of large numbers of particles, the \textbf{classical value} of an observable quantity is given by the expectation of the corresponding Hermitian operator (irrespective of whether the ket is an eigenket).\\
\\
This is weaker than the projection postulate, which requires an eigenket (in which the value is trivially given by the expectation). The reason for this is seen in \cite{Francis}, in which it will be found that the classical electromagnetic field, $ A(x) $, is given by the expectation of the photon field operator.

\section{Quantum covariance}
\label{Quantum Covariance}
\subsection{The Schr\"{o}dinger equation}\label{sec5.1}
The inner product allows us to calculate probabilities for the outcome of a measurement provided that we know the ket describing hypothetical measurement at the time of measurement. This is only useful if we can calculate the ket at any time, $t$, from a known previous measurement result. Hilbert space refers to measurement at time, $t$, so that $|f(t)\rangle \in \mathbb{H}(t)$, where $t$ is a parameter and we isomorphically identify $\mathbb{H}(t) = \mathbb{H}$ for all $t$. The position ket $|x\rangle$ at time $x^0 = t$ will be denoted by $|t,x\rangle$. Since $\mathbb{H}$ has a finite basis, it is required to review the arguments for the Schr\"{o}dinger equation.\\
\\
\textbf{Postulate:} If at time $t_0$ the ket is $|f(t_0)\rangle$, then the ket at time $t$ is given by the \textbf{time evolution operator}, $U(t, t_0): \mathbb{H} \rightarrow \mathbb{H}$, such that $|f(t)\rangle = U(t,t_0)|f(t_0)\rangle$.\\

If the ket at time $t_0$ was either $|f(t_0)\rangle$ or $|g(t_0)\rangle$, then it will evolve into either $|f(t)\rangle$ or $|g(t)\rangle$ at time $t$. Any weighting in \textsc{or} will be preserved. So, $ U $ is linear 
\begin{equation}\label{Eq5.1.1}
U(t,t_0)(a|f(t_0)\rangle + b|g(t_0)\rangle) = aU(t,t_0)|f(t_0)\rangle + bU(t,t_0)|g(t_0)\rangle.
\end{equation}

Irrespective of whether a model of discrete particles might appear continuous on the large scale, the evolution of kets is expected to be continuous because kets are not physical states of matter, but are rather probabilistic statements about what might happen in measurement, given current information. Probabilities describe our ideas concerning the likelihood of events. Whether or not reality is fundamentally discrete, probability is properly described on a mathematical continuum. Between measurements there is no change in information. Then the result of the calculation of probability is not affected by the time at which it is calculated. Since phase is arbitrary, we may choose it to be continuous. So, time evolution is modelled by a continuous operator valued function of time, $ U $. 

Since local laws of physics are always the same, and $U$ does not depend on the ket on which it acts, the form of the evolution operator for a time span $t$, $U(t)=U(t+t_0,t_0)$, does not depend on $t_0$. We require that the evolution in a span $t_1 + t_2$ is the same as the evolution in $t_1$ followed by the evolution in $t_2$, and is also equal to the evolution in $t_2$ followed by the evolution in $t_1$, $U(t_2)U(t_1) = U(t_2+t_1) = U(t_1)U(t_2)$. In zero time span, there is no evolution. So, $U(0)$ does not change the ket; $U(0)=1$. Using negative $t$ reverses time evolution (put $t = t_1 = -t_2$); $U(-t)=U(t)^{-1}$.

Since kets can be chosen to be normalised we may require that $U$ conserves the norm, i.e. for all $|g\rangle$, $\langle g|U^{\dagger} U|g\rangle = |U|g\rangle |^2 = ||g\rangle |^2 = \langle g|g \rangle$. This is sufficient to show that $U$ is unitary (appendix A). Thus the conditions of Stone's theorem \cite{Stone} (appendix B) are satisfied and we have that there exists a Hermitian operator $H$, the Hamiltonian, such that $\dot{U}(t)=-iHU(t)$. This has solution $U(t)=e^{-iHt}$. The Schr\"{o}dinger equation and Newton's first law ($H=E = \text{const}$) follow immediately. $E$ is identified with energy and $m$ with mass.

In a general problem in quantum theory, an initial condition is described by a ket $|f \rangle$ with momentum space wave function $\langle p|f \rangle$, and such that the discrete position function is uniquely embedded into the smooth wave function on  $\mathbb{R}^3$, \eqref{Eq:3.2.4}. Solving the Schr\"{o}dinger equation extends the wave function to $\mathbb{R}^4$, \eqref{Eq:3.3.1}. Then the position function at any time, and in any discrete coordinate system is found restricting to discrete values. Thus we do not require the existence of a physical continuum to define quantum theory using smooth wave functions.

\subsection{Quantum covariance}
\label{sec5.2}
If time and position are not properties of prior space or space-time, but only of relationships found in matter, then it follows that the fundamental properties of elementary particles have no dependency on time or position. This is expressed in the principle that, \textit{the fundamental behaviour of matter is always and everywhere the same}. Incorporated in this law is the notion that local, physically realised, coordinate systems may always be established by an observer in the same way. From this we may infer the general principle of relativity, \textit{local laws of physics are the same irrespective of the coordinate system which a particular observer uses to quantify them}. In classical physics, laws which are the same in all coordinate systems are most easily expressed in terms of invariants, known as tensors. Then the most directly applicable form of the principle of general relativity is the principle of general covariance, \textit{the equations of physics have tensorial form}. 

General covariance applies to classical vector quantities under the assumption that they are unchanged by measurement. But in quantum mechanics measured values arise from the action of the apparatus on the quantum system, creating an eigenket of the corresponding observable operator and we cannot generally assume the existence of a tensor independent of measurement. In practice a change of reference frame necessitates a change of apparatus (either by accelerating the apparatus or by switching to a different apparatus). A lattice describes possible values taken from measurement by a particular apparatus. Eigenkets of displacement are determined by this lattice, i.e. by the properties and resolution of a particular measuring apparatus. So, in general, eigenkets in one frame are not simultaneously eigenkets of a corresponding observable in another frame using another apparatus (c.f. non-commutative geometry, Connes \cite{Connes}
). For the same reason classical tensor quantities do not, in general, correspond to tensor observables. 

The broad meaning of covariance is that it refers to something which varies with something else, so as to preserve certain mathematical relations. If covariance is not now to be interpreted as manifest covariance or general covariance as applicable to the components of classical vectors, then a new form of covariance, \textit{quantum covariance}, is required to express the principle of general relativity, that local laws of physics are the same in all reference frames. Quantum covariance will mean that local laws of physics have the same form in any reference frame but not that the same physical process may be described identically in different reference frames, since the reference frame, i.e. the choice of apparatus, can affect both the process under study and the description of that process. Since coordinates are determined by physical measurement which has finite resolution, under transformation of the coordinate system (passive Lorentz transformation) there is also a change of basis for Hilbert space. Quantum covariance observes that, since the choice of basis is arbitrary and observer dependent, and since Hilbert space contains a continuum of kets $|x\rangle$ for $x \in \mathbb{R}^3$, any breaking of manifest covariance by the choice of basis is irrelevant. \\
\\
\textbf{Postulate:} \textbf{Quantum covariance} will mean that the wave function, \eqref{Eq:3.3.1}, is defined on a continuum, while the inner product is discrete, and that, in a change of reference frame, the lattice and inner product appropriate to one reference frame are replaced with the lattice and inner product of another.\\

Thus, from an initial position function defined on $\mathrm{C}$, the position function at any time is given by
\begin{equation}\label{Eq5.2.1}
\langle x|f \rangle = f(x)|_{\mathrm{S}},
\end{equation}
and if, in a change of reference frame, the space-time coordinate system $\mathrm{S}$ is replaced by $\mathrm{S'}$, the new position function is given by
\begin{equation}\label{Eq5.2.2}
\langle x|f \rangle = f(x)|_{\mathrm{S'}}.
\end{equation}
We have seen that the consistency of quantum covariance is ensured if the support of $\langle p|q \rangle$ is bounded as described in section \ref{sec3.3}. 

The general form of a linear operator, $O$ on $\mathbb{H}$, is, for some complex valued function $O(x,y)$,
\begin{equation}\label{Eq5.2.3}
O = \chi^3 \sum_{x,y \in \mathrm{D}} |x\rangle O(x,y) \langle y|.
\end{equation}
According to quantum covariance, this expression has an invariant form under a change of reference frame (this has important implications for the definition of quantum fields and is shown in \cite{Francis}). The invariance of operators under rotations is perhaps at first a little surprising, particularly when one considers the presumed importance of manifest covariance in axiomatic quantum field theory. It may be clarified a little with a nautical analogy. On a boat the directions fore, aft, port and starboard are invariant because they are defined with respect to the boat. Similarly operators are necessarily defined with respect to chosen reference matter and have an invariant form with respect to reference matter.

\section{Discussion}
\label{Discussion}
\subsection{The measurement problem}
\label{sec6.1}
It has been seen that the principle of superposition is logical disjunction in a formal language describing hypothetical measurement results in the subjunctive mood, and constructed to give probabilistic results for actual measurements. The Schr\"{o}dinger equation has been shown from the requirements of the probability interpretation, by way of unitarity and Stone's theorem, and is an abstract device which does not determine the motion of a mechanistic or material wave, and which does not depend on the physical metric. Thus, the equations of wave mechanics, and hence also quantum interference effects, arise from the mathematical structure of Hilbert space and the requirement that the probability of a measurement result given an initial condition does not change depending on the time when it is calculated (appendix A).

The inherent conflict between determinist wave motion and probabilistic collapse has come to be known as the measurement problem. The interpretation used in this paper can be classed as an information theoretic interpretation. Information theoretic interpretations have their roots in the original discussions between Bohr, Heisenberg, and others, which led to the Copenhagen interpretation, but they discard the notion of complementarity. The wave function is not conceived as describing a fundamental property of matter, but rather it describes what we can say about measurement. \textit{``What we observe is not nature itself, but nature exposed to our method of questioning''} (Heisenberg \cite{Heisenberg})
. It does not describe a physical wave, but is simply a way of calculating the probability for the outcome of an experiment. Information theoretic interpretations invert the measurement problem. Collapse is simply the change in a probability once the outcome of a measurement is known, but wave evolution requires explanation.

The problem with information theoretic interpretations has been that they fall short of being complete interpretations of nature. If the wave function describes what we can know about reality, not reality itself, then we are lacking a description of the underlying physics. We must explain why the laws of quantum mechanics yield correct probabilities and the reason that why evolution obeys the laws of wave mechanics. Here the underlying description is one of particles, but Hilbert space is \textit{defined} so as to yield probabilities. Wave evolution follows from the probability interpretation via Stone's theorem, and is determined by the mathematical requirements of probabilities irrespective of physical mechanism. This shows that quantum theory describes correlations rather than correlata, but does not show that correlata do not exist. Rather, the laws of quantum theory reflect Kant's transcendental idealism, and Plato's allegory of the cave, according to which an ultimate reality exists but is not perceived directly by us, and has a very different fundamental character from that which we do perceive.

\subsection{Locality and causality}
\label{sec6.2}
It is often suggested that the implication of Bell's theorem \cite{Bell} is that, if quantum mechanics is correct, we must sacrifice at least one of locality, causality, and realism. Since physics makes no sense without realism, it seems we must have a problem with either locality or causality, or both. However, Bell's inequality does not directly refer to quantum systems, but rather to classical systems in which the unknowns can be described by hidden parameters. Strictly it does not say that quantum mechanics is non-local, but rather that a theory which reproduces the results of quantum mechanics and in which the unknowns can be described by classical local hidden variables would have to allow either instantaneous propagation or retrocausality.

Quantum theory gives predictions of probabilities for the results of measurements. An alteration to the setting of Alice's instrument does not affect the probability of the result of Bob's measurement. Faster than light signaling is not possible. Only when the results of Alice's and Bob's measurements are brought together, at some later time, does it become possible to ascertain a correlation which cannot be explained by a classical theory. Nevertheless the high correlation predicted by quantum theory creates the \textit{appearance of} a faster than light effect, and this requires explanation.

In quantum electrodynamics spin is an essential feature of the relativistic treatment of particle wave functions, and is intrinsic to the interactions between particles. If space-time is emergent, then spin should be seen as fundamental to its underlying structure. With emergent space-time, the notion of distance between two particles can only be said to hold when the particles exist in space-time, that is to say when there are sufficient interactions between the particles and other matter to establish space-time properties for the particles. This has not happened at the time of Alice's and Bob's measurements in the Bell tests, but it has happened when Alice and Bob get together and determine the correlation. There can be no exchange of photons between the immediate environments of Alice and Bob at the time of their measurements, because this would require that photons travel faster than the speed of light. Therefore, while Alice and Bob each observe space-time structure in their immediate environment, the structure connecting those two regions is not yet complete. At the time when Alice and Bob bring their measurement results together, there will have been many more billions of interactions exchanging photons, and a single space-time structure containing the regions of space-time in which Alice and Bob carry out their measurements can be said to exist. Entanglement is then understood as meaning that space-time relationships have not yet emerged from the interaction between particles and other matter.

The central ingredient of Bell's theorem is the factorisation of independent probabilities using classical probability theory. Specifically, it is assumed that if two variables, $A(a,\lambda)=\pm 1$ and $B(b,\lambda)=\pm 1$, to be measured independently with an assumed spacelike separation, where $a$ and $b$ are unit vectors in directions chosen by Alice and Bob, and $\lambda$ is a hidden variable, then the joint probability can be factorised:
\begin{equation}\label{Eq6.2.1}
P(AB|a,b,\lambda) = P( A|a,\lambda)P (B| b,\lambda).
\end{equation}
However, if we understand probability theory in a modern Bayesian context, then \eqref{Eq6.2.1} expresses a state of \textit{knowledge} about the results of the two measurements. In fact there can be no \textit{simultaneous} knowledge of two events with spacelike separation, and \eqref{Eq6.2.1} is strictly meaningless at the time of the measurements. Later it becomes possible to bring the measurement results together and \eqref{Eq6.2.1} is violated according to the laws of quantum mechanics, but it is not necessary to postulate any superluminal effect because there is common cause and because at this later time space-time has emerged from non-local processes.

In a theory of emergent space-time, Bell's theorem is not an issue. space-time is determined by the configuration of matter. The detail configuration of matter at the level of individual electrons and photons is not known, and cannot be determined. Configuration is non-local, and escapes the constraint of Bell's theorem. We can only express $P(AB|a,b,\lambda)$ when the backward light cone contains both Alice's and Bob's measurements. Since their measurements have common cause, and the unknowns are contained in non-local configuration, we cannot factorise probabilities as in \eqref{Eq6.2.1}. We thus do not have to sacrifice either locality or causality as fundamental principles, but we do have to dismiss naive statements of locality and causality based on an assumption of background space-time. It is necessary to restate locality and causality in a relationist context:\\
\\
\textbf{Definition:} \textbf{Locality.} A particle is in contact with another when it interacts with it. A particle can be considered to be in a neighbourhood of another if, in principle, a photon can be emitted by the particle and absorbed by the other, and then a second photon emitted by the second particle and absorbed by the first within a small proper time period of the first particle.\\
\\
This relationist definition reflects the locality condition in qed (also called microcausality), as well as the relativistic definition of the metric by the radar method, and it allows that entangled particles in Bell's theorem are separated, in accordance with our intuitive ideas.\\
\\
\textbf{Definition:} \textbf{Causality.} There is a causal relation between two measurements if the outcome of one measurement alters the probability of the outcome the other.\\
\\
By this definition there is no causal relationship between the measurements of the entangled particles by Alice and Bob; the measurement of one particle does not alter the probabilities for the results of measurement of the other, for the reason that at the time of his measurement Alice cannot know the result of Bob's measurement. Only when the two experimenters get together and compare results do they find a correlation. This can only be done at a later time, showing that the correlation is causally related to the measurements, but not that the measurements are causally related to each other.

\subsection{Delayed choice experiments}
\label{sec6.3}
In 1978 J. A. Wheeler \cite{Wheeler} recognized that, according to the laws of quantum mechanics, in a Young's slits experiment it should be possible in principle to ``\ldots choose whether the photon (or electron) \textit{shall have} come through both of the slits, or only one of them, after it has \textsl{already} transversed the doubly slit screen'' (Wheeler's italics). A number of delayed choice experiments have now been performed, such as the delayed choice quantum eraser by Kim et al. \cite{Kim}
, and, in the purest form (using individual photons) by Jacques et al. \cite{Jacques}
. The experiments confirm the prediction of quantum mechanics that behaviour at the slits can \textit{apparently} be determined after the particle passes through them.

Although this result is strongly suggestive of retrocausality, it is not necessary to invoke a notion of retrocausality to either to explain delayed choice experiments or to understand the correlation in Bell tests. If space-time is an emergent quantity, it can only be used to describe the behaviour of matter when sufficient contact relationships (interactions) exist in the process under study. We can only say which slit a particle comes through if the particle has sufficient contact relationships with other matter to define position with respect to the slits. An electron passing through the slits does not interact with the environment, and does not participate in the structure of space-time created by other matter in the environment. It therefore cannot be said that the electron passes through either slit. In a delayed choice measurement, spacial relationships are not determined at the time at which a particle passes through the slits, but only later, when they become established through interactions with matter, including interactions taking place after the decision on whether to perform a ``which slit'' measurement. The path of the electron is a post hoc construction contingent upon eventual measurement. Thus, in this scenario there is no retrocausality in the behaviour of matter, but there is a retroactive notion of space-time.

\subsection{Quantum field theory} \label{sec6.4} 
A development of quantum field theory from the foundations described here is the subject of \cite{Francis}. Using Fock space constructed from a finite dimensional single particle Hilbert space, creation and annihilation operators, and hence also quantum fields, are operator valued functions, not operator valued distributions as is usually the case. There is therefore no mathematical problem with the equal point multiplication. Conceptually, reality is described as graphs (Feynman diagrams) showing time lines of electrons where the configuration of the interactions with photons is not known; all possibilities must be summed under the identification of addition with 
\textsc{or}. divergences in loop integrals 

In standard treatments of qed, Feynman diagrams are regarded as aids to calculation, not descriptions of underlying structure. By contrast, here the perturbation expansion can be interpreted directly as a quantum-logical statement, meaning that any number of interactions might be found taking place at any time and any position if we were to do a measurement. The sums in the expansion simply represent \textsc{or} between possibilities. The interaction Hamiltonian describes the possibility that an interaction \textit{might} be \textit{anywhere}, not some form of ``matter field'' which is, in some sense, everywhere. Similarly, Feynman's path integral, or ``sum over all paths'' has as natural interpretation as a logical \textsc{or} between the possible paths that might be detected if an experiment could be done to trace the path (\textit{not} that a particle passes through all paths in spacetime; e.g. Feynman \cite{Feynman4}
).

The meaning of the perturbation expansion is that, since we cannot say how many interactions take place in any given physical process, we sum over possibilities. In a particle interpretation, Feynman diagrams give a pictorial representation of the fundamental structure of matter. We cannot say what the precise configuration of particle interactions in any given instance, but we represent each possible configuration as a graph and sum over the possibilities, using the interpretation of sum as logical disjunction. Only the topology of lines and vertices is relevant. The paper on which the diagram is drawn has no meaning. Spacetime structure does not appear in Feynman diagrams, except in so far as energy-momentum is four dimensional. Thus Feynman diagrams describe the fundamental structure of a particulate relational model in which only particles exist and in which other properties, including spacetime geometry, emerge from interactions between particles.

\section{Conclusions}
\label{Conclusions}
It has been established that formal conditional clauses about hypothetical measurement results have the natural structure of a finite dimensional Hilbert space in which the inner product can be understood as giving complex truth values for statements in the subjunctive mood. Coefficients are constrained by probabilities which apply when hypothetical measurements are replaced by actual measurements and the subjunctive mood is replaced by a factual conditional. Thus the formal mathematical structure of quantum mechanics can be abstracted from ordinary language about measurement results. This interpretation clarifies the view of von Neumann that quantum logic is a language which tells us what can be known from measurement by providing explicit statements in English, corresponding to the mathematical symbolism. 

Quantum mechanics has been formulated here in terms of discrete measurement results at finite level of accuracy in a manner which does not depend on an assumption of a substantive, or background, space-time continuum. It has been shown that, for any coordinate system, discrete position functions are uniquely embedded into smooth wave functions in such a way that differential operators are defined. Because the range and resolution of real measurement is always finite, only a formulation using a discrete basis of measurement results from specific apparatus can be justified from strict empiricism, but the continuum equations remove the dependency on specific measurement apparatus because they contain embedded within them the solutions for all discrete coordinate systems possible in principal or in practice.

The Schr\"{o}dinger equation has been shown from the requirements of the probability interpretation. Wave functions are directly related to probabilities and do not describe an objective property of matter. Instantaneous collapse of the wave function is merely the collapse of a conditional probability when the condition becomes known. Thus Schr\"{o}dinger's cat is not an objective superposition of quantum states, but simply a probabilistic statement that if the box were to be opened there would be a 50-50 probability of finding the cat alive or dead.

Correlations in Bell tests and the results of delayed choice experiments are seen as arising because space-time is an emergent property, seen in measurement but not in the fundamental structures of matter. Experimental results depend on the configuration of matter on a scale below that for which we can have precise knowledge. Since configuration is a non-local property, there is no reason to postulate either retrocausality or non-local effects in the fundamental components of matter.\\
\section*{Appendices}
\begin{appendix}

\section{Unitarity of U}
\label{Ap1}
In the absence of further information, the result of the calculation of probability of a measurement result $ g $ at time $ t_2 $ given an initial condition $ f $ at time $ t_1 $ is not affected by the time at which it is calculated (parameter time for Hilbert space). Since kets can be chosen to be normalised we may require that $U$ conserves the norm, i.e., for all $|g\rangle \in \mathbb{H}$, $\langle g | U^{\dagger}U |g \rangle = \langle g | g \rangle$. Applying this to $|g\rangle + |f\rangle$,
\begin{equation}\label{AEq1}
		(\langle g | + \langle f |)U^{\dagger}U (|g \rangle + |f \rangle) =(\langle g | + \langle f |) (|g \rangle + |f \rangle).
\end{equation}
By linearity of $U$,
\begin{equation}\label{AEq2}
		(\langle g |U^{\dagger} + \langle f |U^{\dagger}) (U|g \rangle + U|f \rangle) =(\langle g | + \langle f |) (|g \rangle + |f \rangle).
\end{equation}
By linearity of the inner product,
\begin{equation}\label{AEq3}
\begin{split}
\langle g|U^{\dagger}U |g \rangle + \langle g|U^{\dagger}U |f \rangle + \langle f|U^{\dagger}U |g \rangle + \langle f|U^{\dagger}U |f \rangle \\ = \langle g |g \rangle + \langle g|f \rangle + \langle f |g \rangle + \langle f|f \rangle.
\end{split}
\end{equation}
\begin{equation}\label{AEq4}
\langle g|U^{\dagger}U |f \rangle + \langle f|U^{\dagger}U |g \rangle  =  \langle g|f \rangle + \langle f |g \rangle .
\end{equation}
Similarly, conservation of the norm of $|g\rangle + i |f\rangle$ gives
\begin{equation}\label{AEq5}
\langle g|U^{\dagger}U |f \rangle - \langle f|U^{\dagger}U |g \rangle  =  \langle g|f \rangle - \langle f |g \rangle .
\end{equation}
Combining \eqref{AEq4} and\eqref{AEq5} shows that $U$ is unitary, i.e. for all $|f\rangle,|g\rangle \in \mathbb{H}$,
\begin{equation}\label{AEq6}
\langle g| U^{\dagger}U |f \rangle =\langle g |f \rangle.
\end{equation} 

\section{Stone's theorem}\label{Ap2}
\begin{flushleft}
\textbf{Theorem:} (Marshall Stone \cite{Stone}
. Let $\left\{ U(t)|t \in \mathbb{R} \right\}$ be a set of unitary operators on a Hilbert space, $\mathbb{H}$, $U(t): \mathbb{H} \rightarrow \mathbb{H}$, such that $U(t+s) = U(t)U(s)$ and 
\begin{equation*}
\forall t_0 \in \mathbb{R}, |f\rangle \in \mathbb{H}, \lim_{t \rightarrow t_0} U_t |f\rangle = U_{t_0} |f\rangle
\end{equation*}
\end{flushleft}
then there exists a unique self-adjoint operator $H$ such that $U(t) = e^{-iHt}$.\\
\\
\textit{Proof:} The derivative of $U$ is
\begin{equation}\label{EqB1}
\begin{split}
\dot{U}(t) &= \lim_{dt\rightarrow 0}\frac{U(t+dt)-U(t)}{dt}= \lim_{dt\rightarrow 0}\frac{U(dt)U(t)-U(t)}{dt} \\ &= \left(\lim_{dt\rightarrow 0}\frac{U(dt)-1}{dt}\right)U(t)= U(t)\left(\lim_{dt\rightarrow 0}\frac{U(dt)-1}{dt}\right)
\end{split}
\end{equation}
This prompts the definition of the Hamiltonian operator:\\
\\
\textbf{Definition:} The \textbf{Hamiltonian} $H: \mathbb{H} \rightarrow \mathbb{H}$ is given by
\begin{equation}\label{EqB2}
H=i\left(\lim_{dt\rightarrow 0}\frac{U(dt)-1)}{dt}\right).
\end{equation}
The Hamiltonian has no dependency on $t$. We have
\begin{equation}\label{EqB3}
\dot{U}(t)=-iHU(t) = -iU(t)H.
\end{equation}
So $-iH= U^{\dagger}\dot{U} = \dot{U}U^{\dagger}$. Since $U$ is unitary, for a small time $dt$,
\begin{equation}\label{EqB5}
1 = U^{\dagger}(t+dt)U(t+dt) \approx [U^{\dagger}(t) + \dot{U}^{\dagger}(t)dt][U(t) + \dot{U}(t)dt] 
\end{equation}
Ignoring terms in squares of $dt$, and using $-iH= U^{\dagger}\dot{U}$, $iH^{\dagger}= \dot{U}^{\dagger}U$
\begin{equation}\label{EqB7}
U^{\dagger}(t)U(t) - iH^{\dagger}dt + iHdt \approx 1.
\end{equation}
Using unitarity of $U$, we find that $H$ is Hermitian, $H=H^{\dagger}$. \eqref{EqB3} has solution, \begin{equation}\label{EqB8}
U(t) = e^{-iHt}.
\end{equation}
\\
\textbf{Corollary:} The wave function satisfies the Schr\"{o}dinger equation
\begin{equation}\label{EqB10}
\partial_0 f(t,x) = -iH f(t,x) .
\end{equation}
\textit{Proof:} Differentiate the wave function using \eqref{EqB3},
\begin{equation}\label{EqB11}
\partial_0 f(t,x)= \langle x | \dot{U} |f(0) \rangle = \langle x | - iHU(t) |f(0) \rangle = \langle x | -iH |f(t) \rangle
\end{equation}
\\
\textbf{Corollary:} Newton's first law.\\
\\
\textit{Proof:} After replacing 3-vectors with 4-vectors in \eqref{Eq3.1.2} and imposing the mass shell condition, $E^2 = (p^0)^2 = m^2 + \boldsymbol{p}^2$ for some constant $m$, we find that a plane wave is a solution of the Schr\"{o}dinger equation with  $H=E= \text{const}$. Thus momentum, $p$, does not change in time for a non-interacting particle.
\end{appendix}


\begin{thebibliography}{10}

\bibitem{Varadarajan}
Varadarajan V. S., Bull. Am. Math. Soc., {\bf 83} (1977) 2.
\bibitem{Rovelli1}
Rovelli C., Relational Quantum Mechanics, Int. J. Th. Phys., {\bf 35} (1996) 1637.
\bibitem{Bub}
Bub J., Interpreting the Quantum World (1997) Cambridge University Press.
\bibitem{Dirac}
Dirac P. A. M., Quantum Mechanics (1958) 4th Ed, p.47 Clarendon Press, Oxford.
\bibitem{von Neumann}
von Neumann J., (1955) Mathematical Foundations of Quantum Mechanics, Princeton University Press.
\bibitem{Heisenberg}
Heisenberg W., Physics and Philosophy (1962) Harper \& Row, New York.
\bibitem{Mermin}
Mermin D., What is quantum mechanics trying to tell us?, American Journal of Physics, {\bf 66} (1998) 753-767.
\bibitem{Adami}
Adami C. and Cerf N. J., 
Proc 1st NASA workshop on Quantum Computation and Quantum Communication, 
quant-ph/9509004.
\bibitem{Jauch}
Jauch J.M., 1968, Foundations of Quantum Mechanics, Addison-Wesley, Reading, Massachusetts
\bibitem{Francis}
Francis C. E. H., A construction of full QED using finite dimensional Hilbert space. EJTP {\bf 10}, No. 28, (2013) 27-80
\bibitem{Dieks}
Dieks D., Stud. Hist. Phil. Mod. Phys., {\bf 32} (2001) No 2, 217-241 (and refs cited therein).
\bibitem{Smolin}
Smolin L., The Future of Spin Networks (1997) gr-qc/9702030 (and refs cited therein).
\bibitem{Rovelli2}
Rovelli C., Quantum space-time, What Do We Know? Physics Meets Philosophy at the Planck Scale (2000) ed. C. Callander, N. Nugget, CUP (and refs cited therein).
\bibitem{Poulin}
Poulin D., Int.J.Theor.Phys., {\bf 45} (2006) 1189.
\bibitem{Svozil}
Svozil K., Quantum Logic (1988), Springer, Singapore, and references cited therein.
\bibitem{Birkhoff}
Birkhoff G. and von Neumann J., (1936) Annals of Mathematics, {\bf 37}, No 4.
\bibitem{d'Inverno}
d'Inverno R., Introducing Einstein's Relativity (1992) Clarendon Press, Oxford.
\bibitem{Rindler}
Rindler W., Special Relativity (1966) Oliver \& Boyd, Edinburgh.
\bibitem{Misner}Misner C. W., Thorne K. S., Wheeler J. A., Gravitation, (1973) Freeman, San Francisco.
\bibitem{Rescher}Rescher N., Many-valued Logic (1969), McGraw-Hill, New York.
\bibitem{Lukasiewicz}\L{}ukasiewicz J., 1920, O logice trojwartosciowej, Ruch Filozoficny, 5: 170-171. \bibitem{Jeffreys}
Jeffreys H. 1939. Theory of Probability. 1st ed. Oxford: The Clarendon Press.
\bibitem{Jaynes}
Jaynes E.T., 2003, Probability Theory: The Logic of Science, Cambridge, Cambridge University Press
\bibitem{Zadeh}
Zadeh L.A., 1965, Information and Control, 8, pp338-353
\bibitem{Greisen}
Greisen, K., End to the Cosmic-Ray Spectrum?, Phys. Rev. Let., {\bf 16} (1966) 748-750.
\bibitem{Zatsepin}
Zatsepin G. T., Kuz'min V. A., Upper Limit of the Spectrum of Cosmic Rays, JETPL, {\bf 4} (1966) 78-80.
\bibitem{Stone}
Stone, M. H., Annals of Mathematics {\bf 33} (1932) 643-648. 
\bibitem{Connes}
Connes A., J. Math. Phys., {\bf 41} (2000) 3832-3866.
\bibitem{Bell}
Bell J. S.,
On the Einstein Podolsky Rosen Paradox, 
Physics,  {\bf 70}1 (1964)  195-200
\bibitem{Wheeler}
Wheeler J. A., in Mathematical Foundations of Quantum Theory (1978) ed. Marlow A.R., Academic Press.
\bibitem{Kim}
Kim Y-H., Yu R., Kulik S.P., Shih Y.H., Scully M. O., A Delayed Choice Quantum Eraser. Phys. Rev. Let. {\bf 84} (2000) 1-5. arXiv:quant-ph/9903047.
\bibitem{Jacques}
Jacques V., Wu E., Grosshans F., Treussart F., Grangier P., Aspect A., Rochl J-F., Experimental Realization of Wheeler's Delayed-Choice Gedanken Experiment. Science {\bf 315} (2007) 5814: 966-968.
\bibitem{Feynman4}
Feynman R., QED The Strange theory of light and matter (1985) Princeton University Press.
\end{thebibliography}
\end{document}